# Dissipation of Nonlinear Alfven Waves with Current Sheets in Relativistic Plasmas

Edison Liang
Rice University, Houston, TX 77005-1892

Abstract

We present results from 2.5-dimensional Particle-in-Cell simulations of the interaction of nonlinear Alfven waves with thin current sheets in relativistic plasmas. We find that the Alfven waves cause the current sheet to bend and kink and increase its dissipation. The electrons are eventually heated to form a double Maxwellian, with the hotter Maxwellian caused by the current sheet dissipation and cooler Maxwellian caused by the Alfven turbulence cascade. These results may have important implications for the kinetic dissipation of MHD turbulence in which both nonlinear Alfven waves and current sheets are present, such as turbulence in accretion flows driven by the saturated magnetorotational instability (MRI).

## 1. Introduction

A fundamental problem in astrophysical plasmas is the dissipation of MHD turbulence and the resultant heating and nonthermal energization of electrons. MHD turbulence is ubiquitous in the universe, from accretion disks to the interstellar medium. As the turbulence cascades down to the kinetic level, it dissipates much of its energy via anomalous (collisionless) heating of the electrons and ions (Boyd and Sanderson 1969). Most recent works have focused on nonrelativistic plasmas. Also current sheets are a prevalent feature of MHD turbulence and their role in the turbulence dissipation has been a major unsolved problem. In this paper we present new results on the dissipation of nonlinear shear Alfven turbulence cascade in relativistic plasmas. We focus on the interaction of nonlinear Alfven waves with transverse current sheets. Using the 2.5 D (2-D space, 3-momenta) Particle-in-Cell (PIC Birdsall and Langdon 1991) code Zohar (Langdon and Lasinski 1976) we obtain interesting new results which are absent when only Alfven waves are present without strong current sheets. We find that the electrons are heated to form a double Maxwellian, with the hot component arising from current sheet dissipation and the cooler component arising from wave turbulence cascade.

Our work is partially motivated by recent observations that the MHD turbulence of saturated MRI-driven accretion flows is dominated by large numbers of folded thin current sheets formed by the differential rotational stretching of the azimuthal field $B_\phi$ (Obergaulinger 2008, Gammie et al 2004). At the same time, nonlinear Alfven modes are generated in $B_r$ and $B_z$ by the MRI (Balbus and Hawley 1991, 1992). The interaction of the vertical and radial Alfven modes with the azimuthal current sheets causes the current sheets to warp and kink, enhancing their dissipation (Zenitani and Hoshino 2005). Here we present the preliminary results for sample kinetic simulation results of such a scenario. However, at this point the (explicit) kinetic simulations can only be performed on small scales, while the MHD turbulence structure so far has only been observed on much larger (MHD) scales >> kinetic scales. Provided that the turbulence is truly self-similar (Goldreich and Sridhar 1995), the turbulence and current sheet structure we see on MHD scales should persist all the way down to the kinetic scale. Hence it is still interesting to see what happens to dissipations at the kinetic level, even though the boundary and driving conditions may be somewhat hypothetical.

## 2. PIC simulations

Fig.1 shows a sample problem setup of our 2.5 D PIC simulations. Typically we use a 1024 x 1024 grid ($10^6$ cells) with 16-64 superparticles (numerical representation of a particle) per cell per species, and ion to electron mass ratio $m_i/m_e = 100$. The cell size $\Delta x = \Delta y = 0.5 c/\omega_e$ where $\omega_e$ is the electron plasma frequency. Hence the physical grid has the dimension 512 x 512 $c/\omega_e$, or 51 x 51 $c/\omega_i$ where $\omega_i$ is the ion plasma frequency. Here we show two examples in which the initial eletron temperature $kT_e = 0.25\ m_e c^2$ (nonrelativistic) or $1.5\ m_e c^2$ (relativistic), while the ion temperature $kT_i = 0.25\ m_e c^2$ in both cases. Fig.1 depicts an initial geometry setup. Two opposite current sheets ($J_x$ only) are set up in the top and bottom half of the grid respectively. We have varied the initial configuration of the current sheets in different runs. Their effects are discussed in Sec.4. Counter-propagating nonlinear Alfven waves with $\delta B/B_o = 1$ and narrow bandwidth centered around $\lambda = 256 c/\omega_e$ (=half grid size) are injected in both the x and y directions. Due to the periodic conditions (in both x and y) the counter-propagating waves form a standing wave pattern.

In the absence of current sheets, these nonlinear Alfven standing waves cascade into higher and higher order (shorter wavelength) modes over time, generating wave turbulence (Fig.11). Longitudinal electrostatic modes (Langmuir turbulence Dieckmann et al 2006) are also generated by the colliding Alfven waves via parametric processes. Electrons are heated to temperatures of several $m_e c^2$ by the wave turbulence cascade, and ions are also heated at late times to a comparable temperature (see below). The electron distributions are consistent with a single temperature Maxwellian.

When the $J_x$ current sheets are added, the results become much more complex and interesting. Below we show the detailed results of the relativistic case with $kT_e = 1.5 m_e c^2$, $|B_z| = 10 B_o$, $\Omega_e$(electron gyrofrequency) $= 0.71\omega_e$. Such parameters mimic some of the conditions we find in MRI MHD runs (Hilburn et al 2009). First the current sheets start to warp and kink, driven by the magnetic perturbations of the nonlinear Alfven waves. The folding and warping of the current sheet greatly increases the current sheet surface area (Fig.2). Magnetic dissipation rate is enhanced. Since our 2D geometry forbids x-type reconnection of $B_z$ field (similar to Zenitani and Hoshino 2005), cross field instabilities such as the relativistic drift kink instability (RDKI, Zenitani and Hoshino 2003) are likely important electron heating mechanisms. However, since ours is an e-ion plasma, the lower hybrid drift (LHDI Krall 1971, Gary et al 2008) and other instabilities absent in e+e- plasmas (Zenitani and Hoshino 2005) may also operate here. The relative roles of different instabilities will be studied in future follow-ups papers.

Fig.2 shows the evolution of $B_z$ while Figs.3 and 4 show the evolution of $B_y$ and $B_x$. Initially $B_x$ and $B_y$ form island patterns. Note that unlike $B_z$, $B_x$ and $B_y$ fields can evolve via x-type reconnections. In particular, Fig.4 shows that $J_z$ current sheets must be present and reconnection leads to the formation of continuous $B_x$ domains at late times. Figs.5,6,7 show the evolution of $J_x$, $J_z$ and $J_y$ respectively. $J_z$ and $J_z$ are both induced by the Alfven waves. Note that $J_z$ evolves into a fragmented "atoll" structure at late times. Fig.8 shows the evolution of the charge separation ($n_i - n_e)/n_o$, which is a measure of the induced Langmuir turbulence amplitude. Comparing Fig.8 with Fig.6 shows that charge separation traces the $J_z$ pattern. Figs.9 and 10 show the evolution of electric fields. In Fig.11 we plot the cross-sectional profiles of the Alfven wave amplitudes, showing the cascade towards shorter modes. Fig.12 is a plot of the time evolution of the different energy components. We see that by $t\omega_e = 4000$, approximately 20% of the magnetic energy (mainly $B_z$ field) has been converted into particle energy, mostly hot electrons. Ions also gain energy, but only a fraction of the electron energy. In Fig.13 we show similar plots for the cold electron case ($kT_e=0.25m_ec^2$) for comparison. In both cases we see that the Alfven waves ($B_x$, $B_y$) at first gains energy at the expense of the $B_z$ field, but eventually lose them back to the particles (Fig.13 right panel). Electrostatic energy remains small at all times.

We have examined the evolutions of particle distributions. Fig.14 shows the evolution of the electron distribution for the $kT_e=1.5m_ec^2$ case. At late times we see the formation of a double Maxwellian (plus possibly a steep power law tail, cf. the log-log plot), with the hot component about 5 times hotter than the cooler component. Fig.15 shows the evolution of the ion distribution. Comparing Fig.15 with Fig.14 we see that ion heating proceeds slower than electron heating, due to the ion inertia.

In Fig.16 we compare the electron distribution for 4 different runs, with and without current sheets, and $kT_e = 0.25\ m_ec^2$ vs. $kT_e=1.5m_ec^2$ (discussed above). We see that without current sheets, only a single Maxwellian emerges. But the temperature of this Maxwellian is still lower than the cooler Maxwellian of the current sheet cases. In other words, current sheet dissipation heats electrons much more efficiently than pure Alfvenic wave cascade. This is not surprising, since current sheets imply large amount of magnetic free energy. Once it becomes unstable, it is capable of generating large amplitude electric fields for particle acceleration (Zenitani and Hoshino 2005), whereas pure Alfvenic turbulence accelerates mainly via resonant scattering, (although in the nonlinear limit, the Alfven modes also decay parametrically and transfer some of its energy into electrostatic modes). Details of all these results will be published elsewhere.

3. Astrophysical Applications

Low luminosity black holes (LLBH) such as the one in our Galactic center Sgr A* emit x-rays and radio (Balantyne et al 2007, Melia et al 2001, Melia 2006) which can only be produced by relativistic electrons with temperatures ≥ few MeV (Liu et al 2004, 2006, Yuan 2006). Since the density in the accretion flow is extremely low, coulomb heating via virial ions is negligible (Liu 2005). Currently the favorite heating mechanism is direct collisionless heating of the electrons by the saturated MRI turbulence (Liu 2005). As we see above, pure Alfvenic wave turbulence cascade may not be sufficient to heat the electrons to such high temperatures. But Alfvenic waves plus thin current sheets seem to produce an additional hot component with temperature exceeding 10 MeV. This holds out the hope that thin current sheet dissipation in MRI disks may be the key in producing the ultrahot electrons needed to explain the observed radio to x-ray broadband spectra of Sgr A* and other LLBH's. Moreover, the currently popular synchrotron self-Compton (SSC) model of Sgr A* (Yuan et al 2003, 2004) seems to have difficulty fitting the radio and x-ray spectra (Ohsuga et al 2005) using a single electron temperature. The presence of a 2-Maxwellian electron distribution will make the spectral modeling much more accommodating and flexible (Hilburn et al 2009).

4. Discussions

We have also studied the effects of different initial current sheet configurations by varying the width of the initial current sheet. We have also examined both Harris equilibrium and nonequilibrium (**Curl B** = $4\pi$**J** but uniform density) initial configurations. While such different initial current conditions make some differences in the initial growth rate, the asymptotic state of the plasma and heated electron temperature appear to be rather insensitive to the initial current sheet configurations. Details will be discussed elsewhere.

Figure Captions:

Fig.1 Initial set up of our runs with 1024x1024 doubly periodic grid. In all figures of this paper, x and y coordinates are in units of $c/\omega_e$. A pair of $J_x$ current sheets and $B_z$ fields (in and out of the plane) are set up at t=0. Aflven waves of $\delta B/B_o=1$ and $\lambda$=half grid size are injected from all four boundaries, forming a standing wave pattern. $B_z=10B_o$ in these runs. Figs.2 – 12, 14, 15 all refer to the hot electron case $kT_e=1.5m_ec^2$.

Fig.2 Snapshots of $B_z$ contours at three different times showing the field evolution.

Fig.3 Snapshots of $B_y$ contours at two different times showing the field evolution.

Fig.4 Snapshots of $B_x$ contours at three different times showing the field evolution. Note the sharp field reversals at early times suggesting thin $J_z$ current sheets as seen in Fig.6.

Fig.5 Snapshots of $J_x$ contours at two different times showing the current evolution. The current sheets bend and kink due to pushing by the Alfven modes.

Fig.6 Snapshots of $J_z$ contours at two different times showing the current evolution. At late times $J_z$ exhibits an "atoll" structure.

Fig.7 Snapshots of $J_y$ contours at two different times showing the current evolution.

Fig8 Snapshots of net charge $(n_i-n_e)/n_o$ contours at two different times showing the charge fluctuation evolution. Fig.8 roughly traces the patterns of Fig.6

Fig.9 Snapshot of **E** field contours at early time.

Fig.10 Snapshot of **E** field contours at late time.

Fig.11 Cross-sectional profiles of the Alfven waves at $t\omega_e=4000$, showing the higher frequency modes superposed on the initially smooth profile.

Fig.12 Time evolution of different energy components. $E_{em}$ is total electromagnetic energy.

Fig.13 Time evolution of different energy components for the cool electron case $kT_e = 0.25\ m_ec^2$. Left panel: top to bottom, $E_{Bz}$, $E_e$, $E_i$, $E_{Bxy}$, $E_E$. Righ panel is a blow up of the bottom curves of the left panel: top to bottom, $E_{Bxy}$, $E_{Ex}$, $E_{Ey}$, $E_{Ez}$.

Fig.14 Evolution of the electron distribution in the hot electron case. Left panel: $\log f(\gamma)$ vs. $\log \gamma$. Right panel: $\log f(\gamma)$ vs. linear $\gamma$, showing the double Maxwellian. There is a slight hint of a power-law tail but the statistics cannot justify such claims.

Fig.15 Evolution of the ion distribution in the hot electron case. It shows that ions are heated much more slowly than electrons.

Fig.16 Comparison of the electron distribution evolution for four different runs. Left: no current sheet or $B_z$ field, only Alfven waves. Right: with double current sheets and $|B_z|=10B_o$. Top panels are for the cool electron case and bottom panels are for the hot electron case.

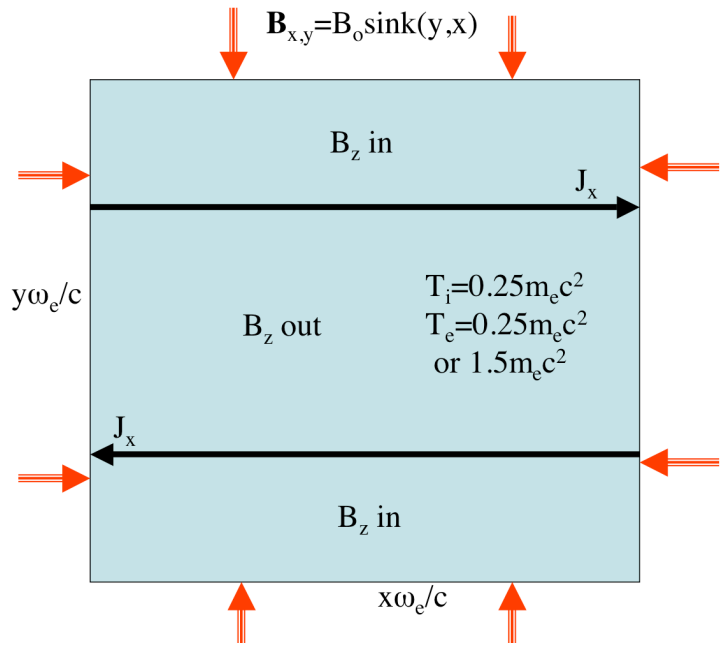

Fig.1

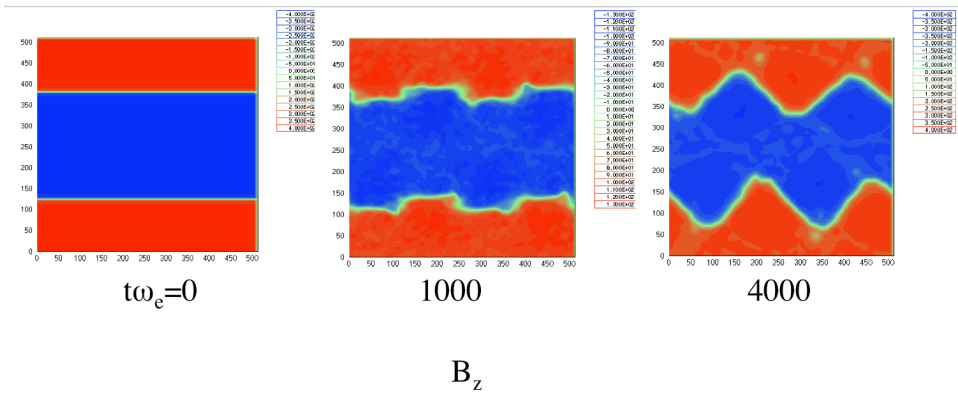

Fig.2

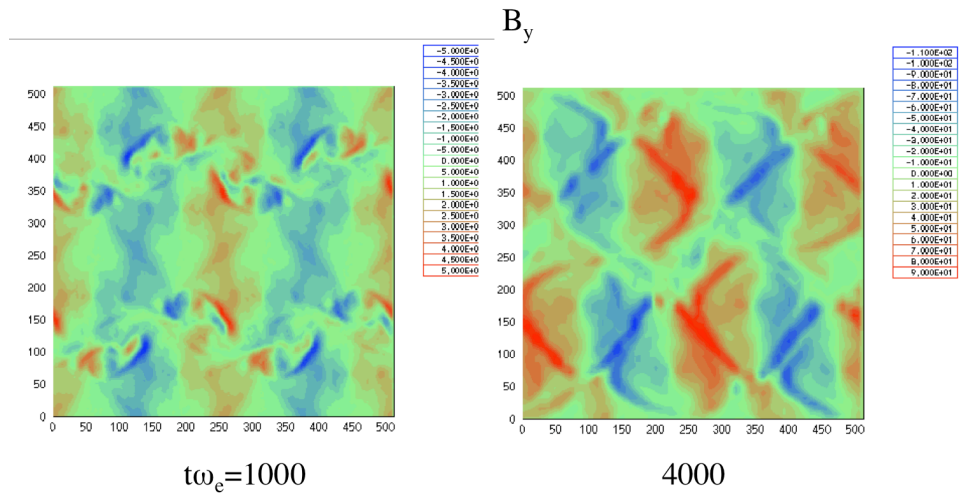

$t\omega_e=1000$            4000

Fig.3

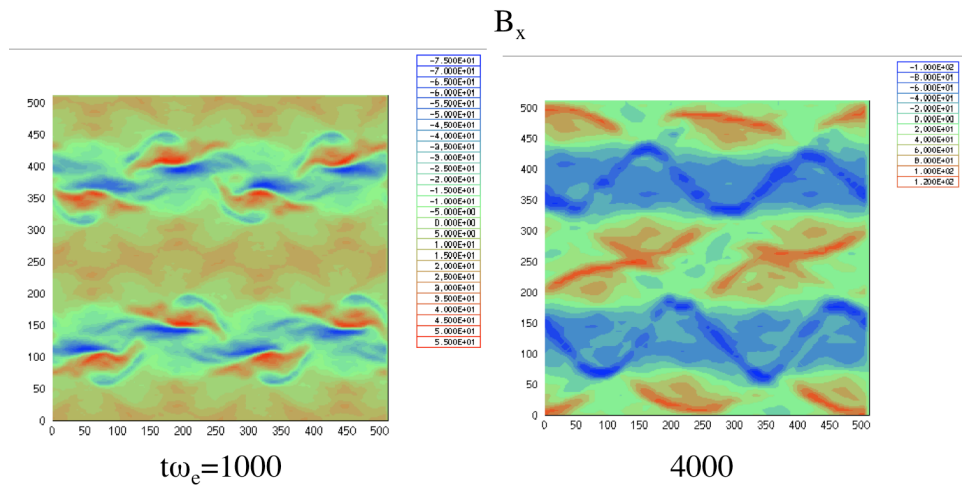

$t\omega_e=1000$            4000

Fig.4

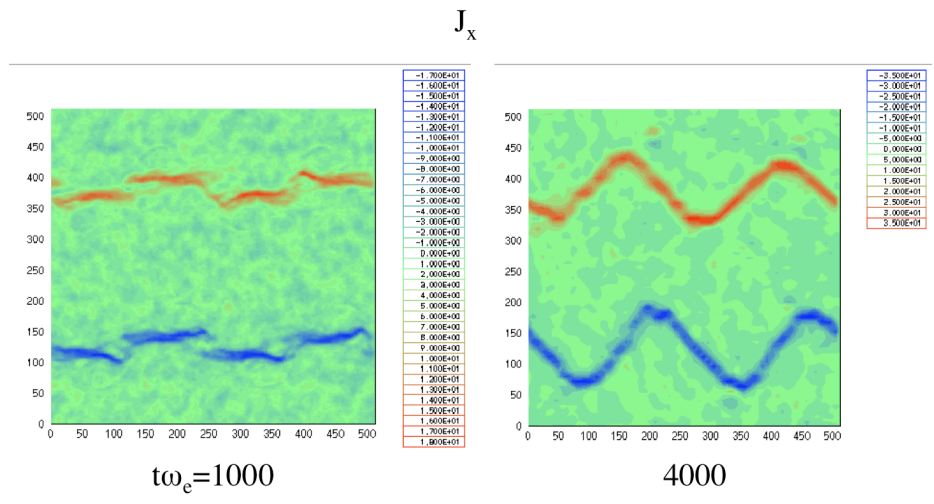

Fig.5

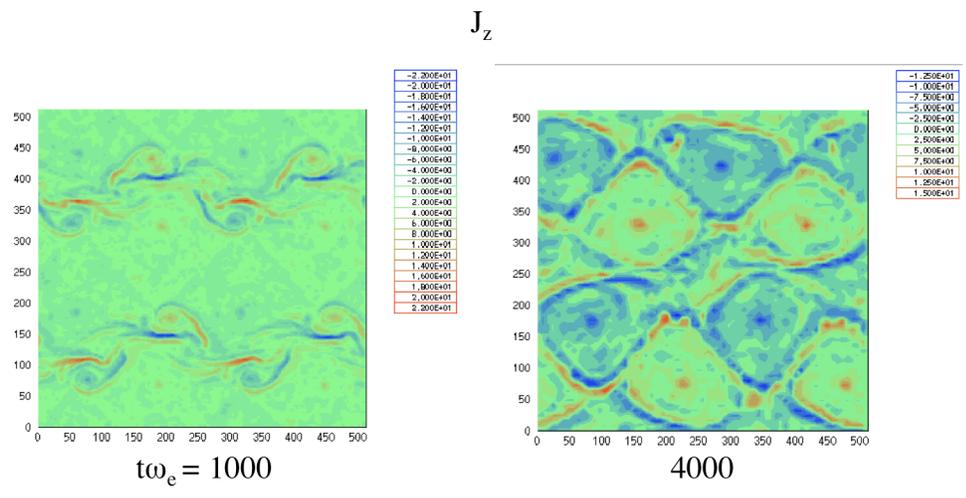

Fig.6

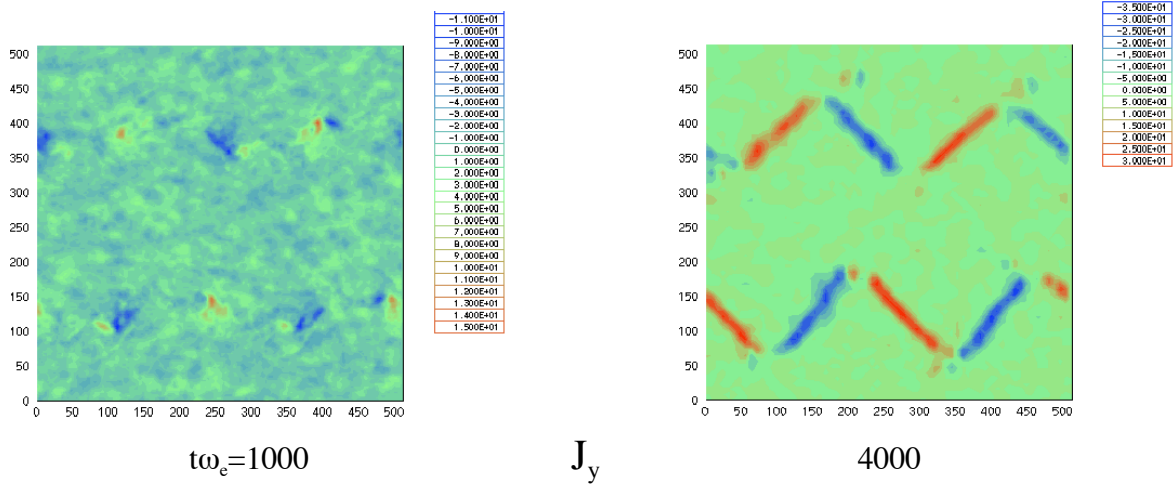

tω_e=1000    $J_y$    4000

Fig.7

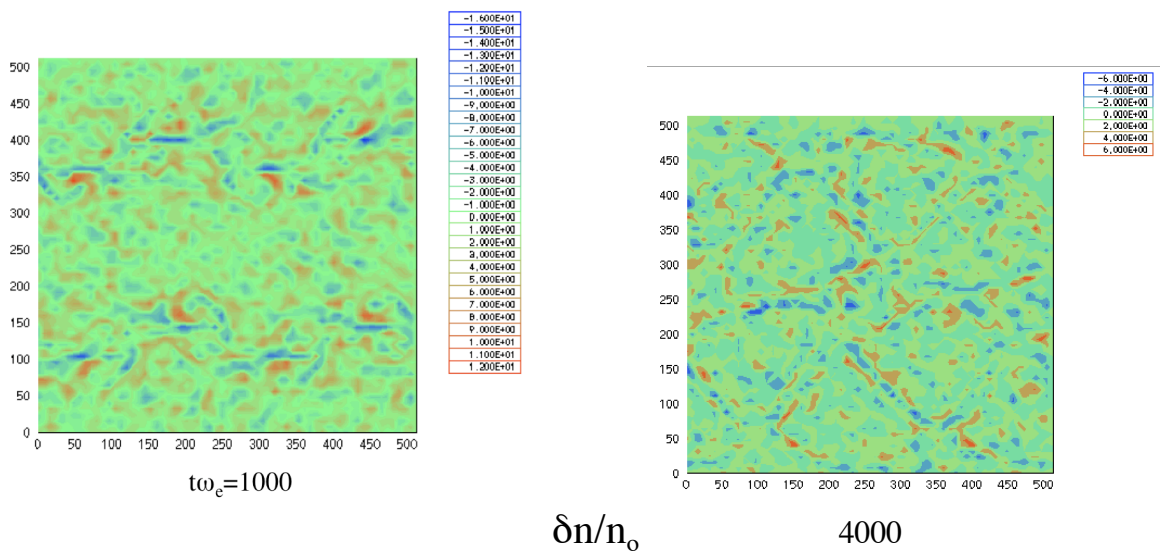

tω_e=1000    $\delta n/n_o$    4000

Fig.8

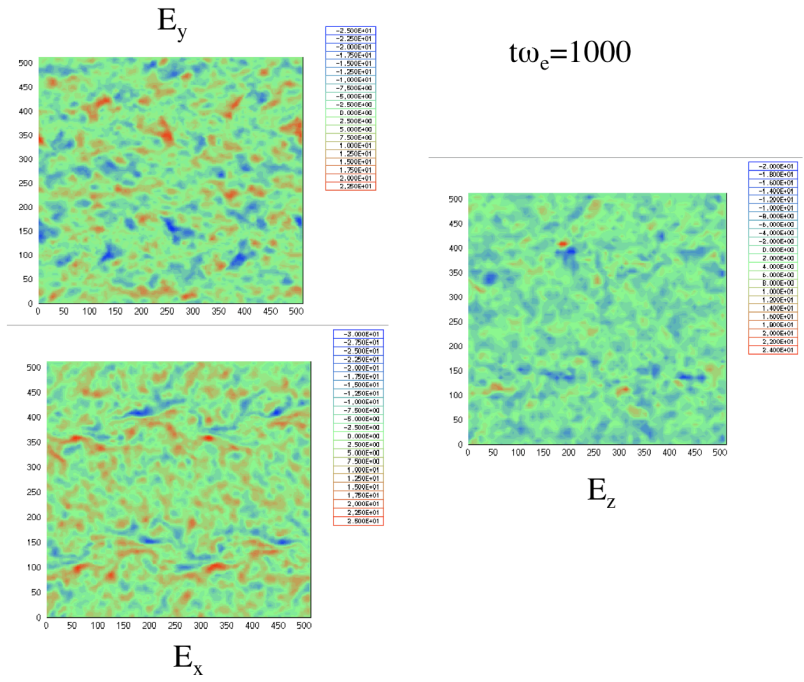

Fig.9

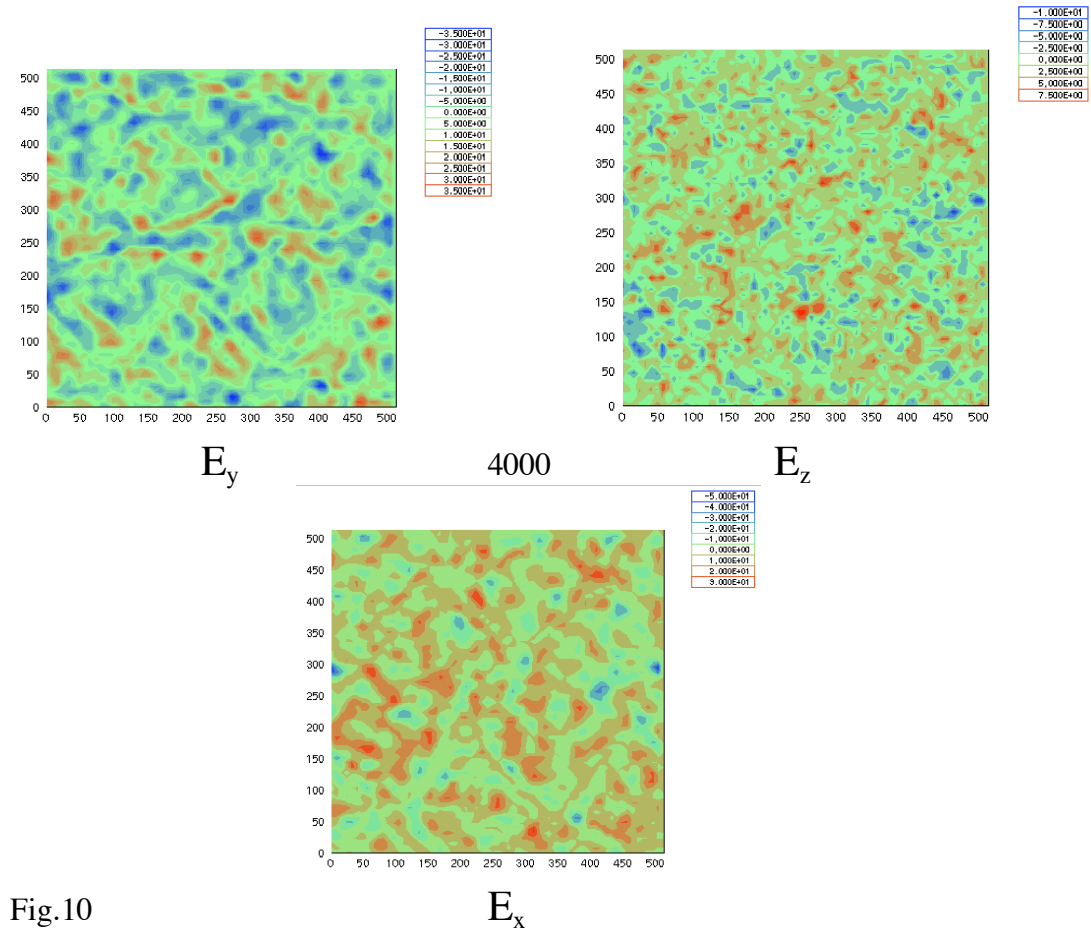

Fig.10

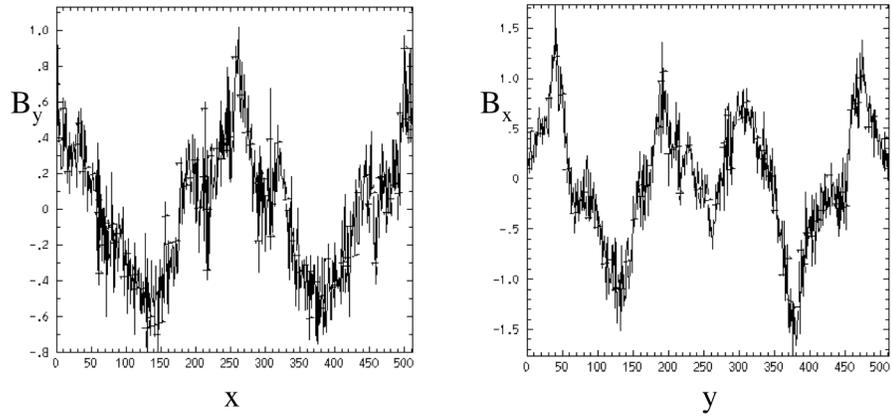

Fig.11

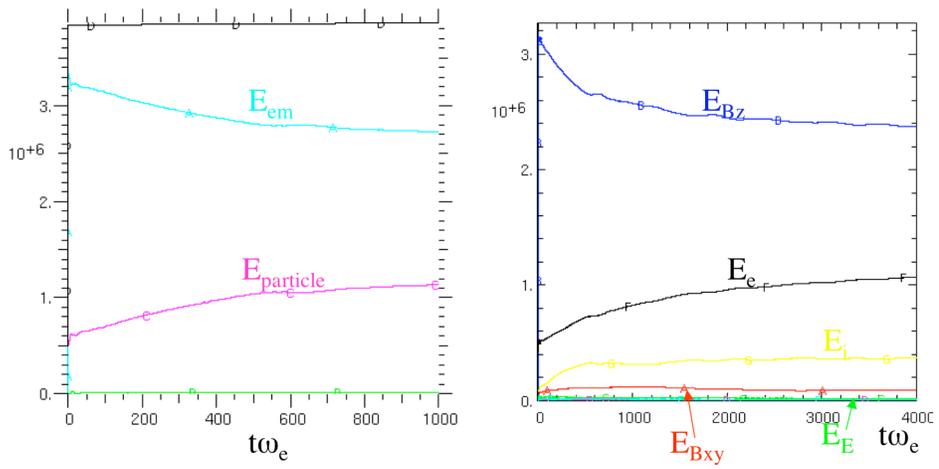

Fig.12

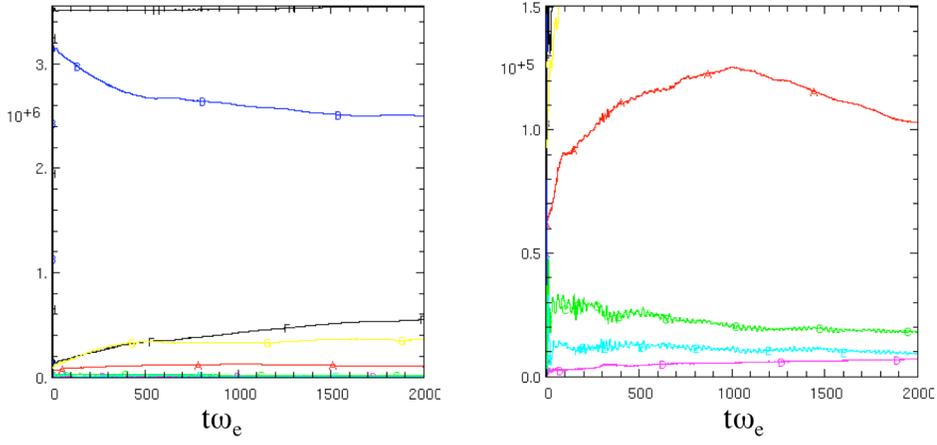

Fig.13

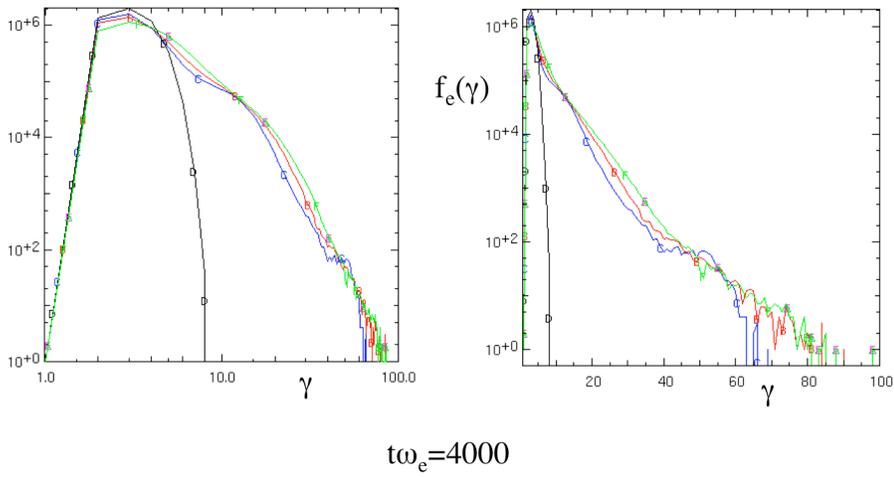

$t\omega_e=4000$

Fig.14

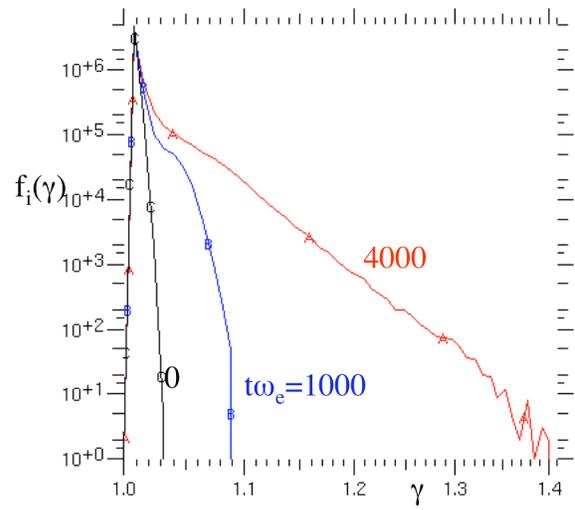

Fig.15

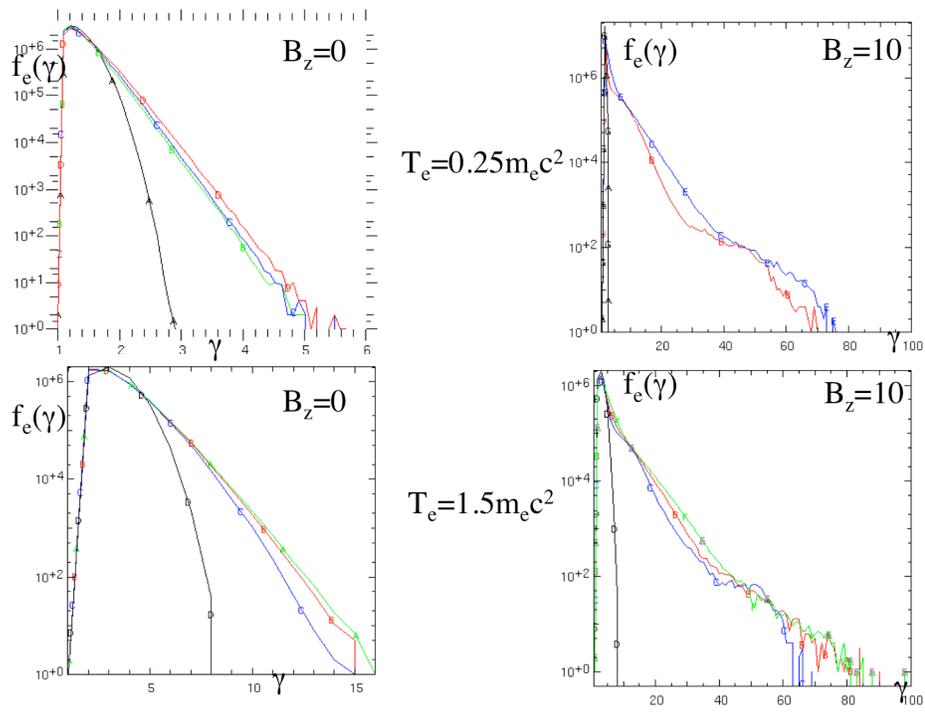

Fig.16